%% file: sample-sigconf.tex
\documentclass[10pt, conference, compsocconf]{IEEEtran}  
\usepackage{algorithm}
\usepackage{algpseudocode}
\usepackage{amsmath}
\usepackage{amssymb}
\usepackage{amsfonts} 
\usepackage{algorithm} 
\usepackage{algpseudocode} 
\usepackage{savesym}
\usepackage{program}
\savesymbol{AND}
\savesymbol{OR}
\savesymbol{NOT}
\savesymbol{TO}
\savesymbol{COMMENT}
\savesymbol{BODY}
\savesymbol{IF}
\savesymbol{ELSE}
\savesymbol{ELSIF}
\savesymbol{FOR}
\savesymbol{WHILE}

\usepackage{scalerel}
\usepackage{pifont}
\DeclareMathOperator*{\minimize}{minimize}

\usepackage{varwidth}

\usepackage{mathtools}
\usepackage{mathptmx}
\usepackage{subfigure}
\usepackage{bbm}

\usepackage{array}
\usepackage{latexsym}
\usepackage{stmaryrd}
\usepackage{color}
\usepackage{epsfig}
\usepackage{boxedminipage}
\usepackage{times}
\usepackage{verbatim}
\usepackage{url}
\usepackage{xspace}

\usepackage{listings}
\usepackage{verbments} 
\usepackage{paralist}
\usepackage{gensymb} 
\usepackage{xcolor}
\usepackage{graphicx}
\usepackage{caption}
\usepackage{listings}
\usepackage{xcolor}

\colorlet{punct}{red!60!black}
\definecolor{background}{HTML}{EEEEEE}
\definecolor{delim}{RGB}{20,105,176}
\colorlet{numb}{magenta!60!black}

\lstdefinelanguage{json}{
    basicstyle=\normalfont\ttfamily,
    numbers=left,
    numberstyle=\scriptsize,
    stepnumber=1,
    numbersep=8pt,
    showstringspaces=false,
    breaklines=true,
    frame=lines,
    backgroundcolor=\color{background},
    literate=
     *{0}{{{\color{numb}0}}}{1}
      {1}{{{\color{numb}1}}}{1}
      {2}{{{\color{numb}2}}}{1}
      {3}{{{\color{numb}3}}}{1}
      {4}{{{\color{numb}4}}}{1}
      {5}{{{\color{numb}5}}}{1}
      {6}{{{\color{numb}6}}}{1}
      {7}{{{\color{numb}7}}}{1}
      {8}{{{\color{numb}8}}}{1}
      {9}{{{\color{numb}9}}}{1}
      {:}{{{\color{punct}{:}}}}{1}
      {,}{{{\color{punct}{,}}}}{1}
      {\{}{{{\color{delim}{\{}}}}{1}
      {\}}{{{\color{delim}{\}}}}}{1}
      {[}{{{\color{delim}{[}}}}{1}
      {]}{{{\color{delim}{]}}}}{1},
}
\IEEEoverridecommandlockouts

\begin{document}
\title{Costless: Optimizing {\bf Cost} of Server{\bf less} Computing through Function Fusion and Placement}

\author{\IEEEauthorblockN{1\textsuperscript{st} Given Name Surname}}
\author{\IEEEauthorblockN{Tarek Elgamal\IEEEauthorrefmark{2},
Atul Sandur\IEEEauthorrefmark{3}, 
Klara Nahrstedt\IEEEauthorrefmark{5}, Gul Agha\IEEEauthorrefmark{6}}
\IEEEauthorblockA{Department of Computer Science, University of Illinois, Urbana-Champaign\\
Email: \IEEEauthorrefmark{2}telgama2@illinois.edu,
\IEEEauthorrefmark{3}sandur2@illinois.edu,
\IEEEauthorrefmark{5}klara@illinois.edu,
\IEEEauthorrefmark{6}agha@illinois.edu}
}

\maketitle


\begin{abstract}
Serverless computing has recently experienced significant adoption by several applications, especially Internet of Things (IoT) applications. In serverless computing, rather than deploying and managing dedicated virtual machines, users are
able to deploy individual functions, and pay only for the time that their code is actually executing. However, since serverless platforms are relatively new, they have a completely different pricing model that depends on the memory, duration, and the number of executions of a sequence/workflow of functions. In this paper we present an algorithm that optimizes the price of serverless applications in AWS Lambda. We first describe the factors affecting price of serverless applications which include: (1) fusing a sequence of functions, (2) splitting functions across edge and cloud resources, and (3) allocating the memory for each function. We then present an efficient algorithm to explore different function fusion-placement solutions and find the solution that optimizes the application's price while keeping the latency under a certain threshold. Our results on image processing workflows show that the algorithm can find solutions optimizing the price by more than 35\%-57\% with only 5\%-15\% increase in latency. We also show that our algorithm can find non-trivial memory configurations that reduce both latency and price.

\end{abstract}

%
%

\input{intro}
\input{background}

\input{models}
\input{approach}
\input{eval}

\input{related}
\input{conc}


\end{document}

%% file: intro.tex
\section{Introduction}
Serverless computing refers to a new generation of
platform-as-a-service offerings by major cloud providers.The first service offered in this category was Amazon Web Services (AWS) Lambda~\cite{lambda} which was first announced at the end of 2014, and experienced significant adoption in mid to late 2016. All the major cloud service providers now offer similar services, such as Google Cloud Functions, Azure Functions and IBM OpenWhisk. 

In Serverless computing, the cloud provider takes responsibility for receiving client requests and responding to them, and performing task scheduling and operational monitoring. Developers need to only write the code for processing client requests. This is a significant change from the traditional paradigm in which development and operations staff have to explicitly manage their virtual machines. With the aid of serverless technology, rather than continuously-running virtual machines, developers can now deploy ‘functions’ that operate as event handlers, and only pay for CPU time when these functions are executing. 

The pricing model of serverless computing depends on the memory allocated to the functions and the CPU time of executing them. In addition, serverless computing services, such as AWS lambda, provide a method to create a {\it workflow} of functions, called {\it state machine}. The state machine specifies the order at which lambda functions are invoked such that the output of one function is the input of the next function. AWS lambda charges an additional cost for each transition from one function to another. Therefore, one way to optimize the cost is to {\it fuse} multiple functions together  and rewrite them as one function to avoid paying for the transition cost. However, it is not always ideal to fuse functions when they have different memory requirements. For example, if one function requires 2GB of memory and takes one second to execute and the next one requires 0.5 GB but needs 5 seconds to execute, fusing them requires executing one long function (6 seconds) with at least 2GB of memory which is not cost effective as we will describe in Section~\ref{sec:background}. Therefore, the decision of which functions to fuse is non-trivial problem especially in the presence of large workflows with more than 10 functions such as scientific workflows~\cite{profiling} and machine learning models~\cite{rl}. We refer to the problem of deciding which functions to fuse as the {\it Function Fusion Problem}.

Another challenge in serverless computing is the {\it Function Placement}. In order to match the increasing volume of data coming from Internet of Things (IoT) devices, AWS offers another service in its serverless computing ecosystem, called AWS Greengrass~\cite{green}. AWS Greengrass allows processing data closer to the source where the data is generated instead of sending it across long routes to data centers or clouds. Greengrass supports running functions on edge devices (e.g., Raspberry Pi) that are controlled by the users and provide a tight integration between the user's edge device and the cloud infrastructure owned by Amazon. Users of Greengrass are charged per device rather than per function so no matter how many functions are running on the edge device, the cost is fixed. However, due to the limited compute capacity on such edge devices, the function execution might be significantly slower. A natural question that arises is which functions to place on the resource constrained edge devices in order to optimize the cost without dramatically increasing the latency. We refer to this problem as the {\it Function Placement} problem.

In this paper we address the problems of {\it Function Fusion} and {\it Function Placement} in order to optimize the cost of deploying serverless applications in AWS lambda and we make the following technical contributions:
\begin{itemize}
\item We highlight the different factors that affect the cost of AWS Lambda as a representative serverless computing service.
\item We formulate the problem of optimizing the price and  execution time of serverless applications. We propose two models: (1) price model for AWS Lambda, and (2) execution time model that estimates the response time of the workflow of functions based on their execution and communication costs.

\item We propose an algorithm to explore possible function fusions and placements. We represent the solutions in a structure that we refer to as the {\it Cost Graph} and we formulate the problem as a Constrained Shortest Path problem in which we find the solution with the best latency within a certain budget and vice versa.





\end{itemize}

The rest of the paper is structured as follows. In Section~\ref{sec:background}, we start with the background about serverless computing pricing and we highlight the factors affecting it. In Section~\ref{sec:models},
we define the pricing and execution time models and we formally define the cost optimization problem. In Section~\ref{sec:approach}, we present the novel function-fusion placement algorithm. In Section~\ref{sec:eval}, we evaluate the proposed method and we compare it with the optimal solutions and other heuristics. In section~\ref{sec:related}, we review the related work. Finally, Section~\ref{sec:conc} concludes the paper.


%% file: background.tex
\section{Background and Motivation}\label{sec:background}
\subsection{AWS Lambda Pricing}
Figure~\ref{fig:workflow} shows an image processing state machine (workflow) with five functions. 
Each node in the workflow is an AWS Lambda function and the arrow describes the dependency between functions. 
The workflow starts by detecting the face in the photo and then matches the face against a collection of previously indexed faces. The photo is then resized to be shown as a thumbnail in the smartphone application,  the user's face is indexed in a collection for future matching, and finally the photo's metadata is saved in the user profile.

The price for each lambda function is calculated using 4 factors:
\begin{enumerate}
\item The number of times each function is executed per month (e.g., 1,000,000 executions/month).

\item The memory allocated to the function by application developers. The CPU resources allocated to the function represent an implicit parameter. This parameter value is proportional to the function's allocated memory (i.e., a 256 MB function is automatically allocated twice the CPU speed than a 128 MB function).
\item The duration how long the function runs.

\item The price per 1 GB of memory and 1 second of execution. For AWS Lambda, the price of 1 GB and 1 seconds is 0.00001667\$/GB-s.  
\end{enumerate}
Figure~\ref{fig:workflow} shows the memory allocation and duration for each function. Assuming that the workflow is executed 1,000,000 times, the price of the first function {\it FaceDetection} is calculated as:
\begin{equation}
\begin{split}
Price_{FaceDetection}=1,000,000\ execution\ *\ 512/1024\ GB \\ \ * \ 2\ seconds *\ 0.00001667\$/GB-s  = 16.67\$
\end{split}
\end{equation}

Similarly, the price for the five functions is given as :

\begin{equation}\label{eq:lambda}
\begin{split}
Price_{lambdaFunctions}= 1,000,000 * 0.00001667 [(512/1024\ \\  *\ 2) + \ (128/1024\ * \ 5)\ +\ (128/1024\ *\ 1.5)\ +\ \\  (256/1024\ *\ 0.3)\ +\ (128/1024\ *\ 0.2)]=35\$
\end{split}
\end{equation}


In addition to the lambda functions price, there is an additional price for each transition between functions, referred to as a state transition price. The state transition price is charged by AWS to handle the message passing and coordination between two successive functions.

The workflow in Figure~\ref{fig:workflow} has 6 state transitions defined by the number of arrows. The total states transitions price is:

\begin{equation}\label{eq:tran}
\begin{split}
Price_{transition}= 6\ transitions\ per\ execution\ *\ 1,000,000 \\ executions\ *\ 0.000025\$\ state\ transition\ price = 150\$
\end{split}
\end{equation}


The total price of the entire workflow includes both the lambda function price (Eq.~\ref{eq:lambda}) and the state transition price (Eq.~\ref{eq:tran}) and it is given by:

\begin{equation}\label{eq:workflow}
\begin{split}
Price_{workflow} = Price_{lambdaFunctions} + Price_{transition} =\\ 35 + 150 = 185\$
\end{split}
\end{equation}

As we mentioned earlier, AWS Greengrass is another service that charges a small per-device fee to connect user-owned edge devices securely to Amazon cloud and no matter how many functions are executed on the edge device, the price does not change. The price of one edge device ranges from 0.16\$ to 0.22\$, based on the region. 

\subsection{Factors affecting price of serverless applications}

Based on the pricing model described above, we identify three major factors that are crucial to the pricing of serverless application workflows. Such factors are : (1) Number of State Transitions, (2) Edge vs. Cloud Computation, (3) Memory allocated to each cloud function. 

In the following, we describe each factor in detail and in the rest of the paper we focus on manipulating the first two factors to optimize the price of serverless applications.

\subsubsection{Number of State Transitions}
Building applications from individual components, where each component performs a small function, makes applications easier to scale and change. However, we note that the transition price can sometimes dominate the price of the entire workflow as shown in Eq.~\ref{eq:workflow}. In such cases, there are incentives to reduce the number of state transitions to make the price lower and within a certain budget without affecting the application correctness.

\begin{figure}
  \includegraphics[width=0.48\textwidth]{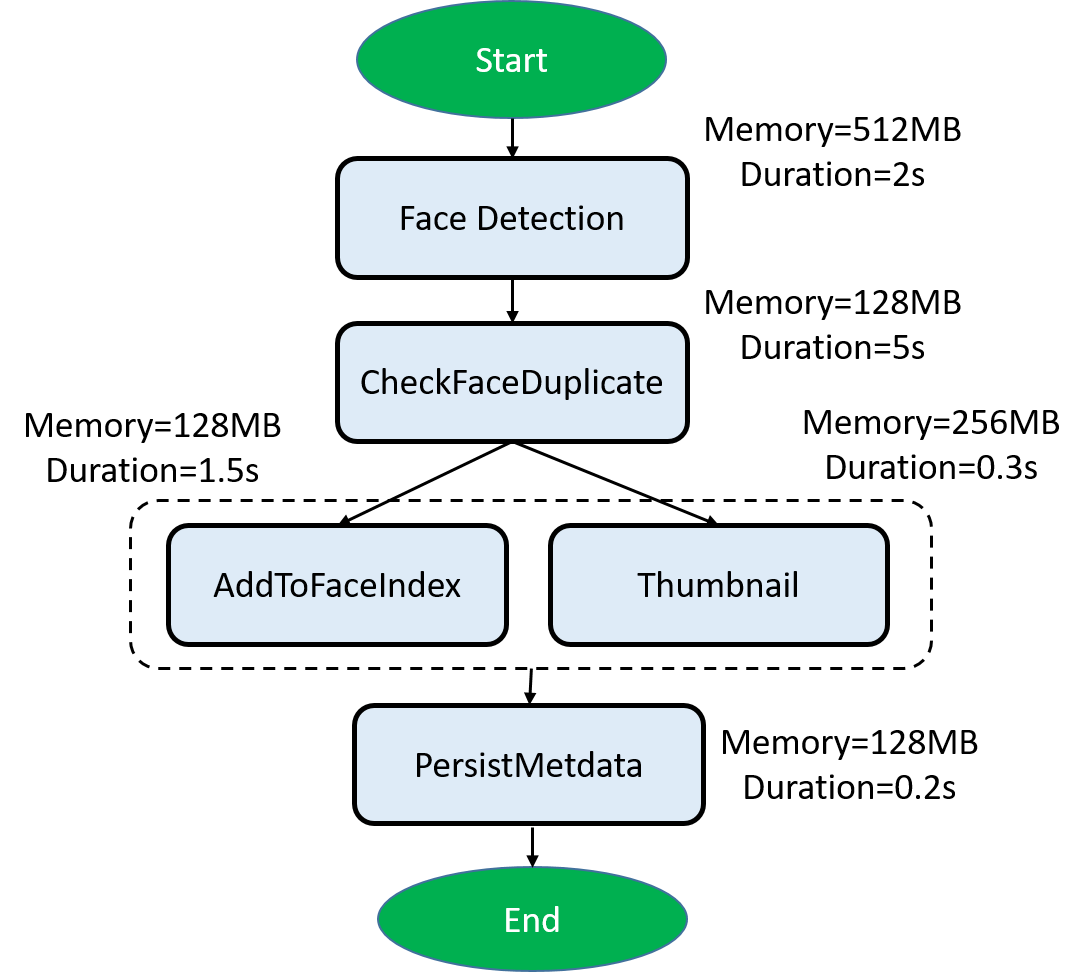}
  \captionof{figure}{Example AWS workflow (state machine)}
  \label{fig:workflow}
\end{figure}

An effective method to reduce the state transitions is to fuse multiple functions to form one bigger function. For example, in Figure~\ref{fig:workflow}, the first two functions, {\it FaceDetection} and {\it CheckFaceDuplicate}, can be fused together to be one function {\it FaceDetAndDup} and remove the state transition between them, which could potentially reduce the cost by 25\$ ($1\ state transition\ *\ 1M\ executions\ *\ 0.000025\$)$.

However, it is not necessarily useful to fuse functions since one function requires 512MB and the other function requires 128MB, and fusing them together will require using at least 512MB for the fused function. Hence, assuming that the second function will still run for a duration of 5 seconds, the cost of the fused functions will then be:

$P_{FaceDetAndDup} = 1,000,000\ *\ 512/1024 \ *\ (5 + 2)\ seconds\ \\ *\ 0.00001667\$ = 58.3\$$

and the previous cost of the non-fused functions used to be :

$P_{FaceDet} + P_{FaceDup} + P_{transition}  = 1,000,000*0.00001667 [(512/1024 \ *\ 2\ seconds)\ +\ (128/1024\ *\ 5)] + 25\$ = 52.3\$$

Hence, in this case the fused function will end up being more costly than original functions but in other cases when both functions have the same memory requirements fusing them can reduce the overall cost . 

Another challenge for the function fusion operation is when trying to fuse parallel functions with their parent. For example, Figure~\ref{fig:workflow} has two parallel functions $AddToFaceIndex$ and $Thumbnail$. If one or both of them are fused with their parent, then fusion will cause two parallel functions to run sequentially and the latency of the entire workflow increases. 

We conclude that it is not trivial to decide which functions to fuse because it could have implications on the price and the latency. In our model we consider both the price and latency of the fused functions and we decide to fuse  functions that can keep the cost under a certain budget while maintaining the best possible latency within the budget constraints. 

\subsubsection{Edge vs. Cloud Computation}
Computing functions on edge devices could be cost effective because no matter how many functions you execute on it, the charge is per-device only and it is relatively cheap ($0.16\$-0.22\$$ per device per month). The edge device typically communicates with the cloud through saving the intermediate data in Amazon's Simple-Cloud-Storage-Service (Amazon S3). There is an additional price for storing the data on S3 but it is also relatively cheap (0.023\$ per GB per month). Due to the limited compute capacity of such edge devices, the function execution might be considerably slower. Therefore, if the application is compute-intensive, it is desirable to place only a subset of the functions on the edge to keep the latency within certain bounds. 

We further note that there is a non-trivial transmission time to send the intermediate data from the edge to the cloud. Therefore, it is desirable to place the functions that reduce the transmission time on the edge device.  In our model we consider both computation and transmission times and we choose the best placements with respect to both price and latency.

\subsubsection{Memory Allocation for each Function}

AWS Lambda allows developers to allocate memory for their function. The CPU resources are not directly configurable because AWS allocates CPU proportional to the allocated memory. For example, AWS allocates twice as much CPU power for a function while going from 128MB to 256MB of memory. However, based on the function implementation, and whether it is compute-intensive or not, if it runs for 4 seconds with 128MB, it may not run for exactly 2 seconds when it is switched to 256 MB. Some functions may run faster than 2 seconds and some may run somewhere between 2 and 4 seconds and after increasing the memory to a certain value, the execution time tends to stabilize because the code does not fully utilize the CPU. 

Tuning the memory for each function separately is a difficult problem and directly impacts the price and the latency of serverless applications. We note that exploring different memory configurations is fundamentally similar to exploring different placements of the function between edge and cloud resources. Intuitively, placing a function on an edge device is fundamentally similar to placing a function on a VM with 128MB or a VM with 512MB. The only effect is that the execution time changes. Hence, in our algorithm we not only explore placing functions on edge devices but we  can also explore placing them on 128MB cloud VM or 256MB cloud VM. For the simplicity of discussion, we focus for the rest of the paper on one edge and one cloud configuration.

%% file: models.tex
\section{Models and Problem Definition}\label{sec:models}

\subsection{Resource model:} We consider two components of the serverless computing platforms one edge $E$ and one cloud $C$. The edge resource is a device close to the data sources (i.e., IoT devices) and owned by the user. An example of such edge devices are Raspberry Pi, and personal desktop. The edge devices use Greengrass core software that provides a tight integration between the edge device and AWS cloud infrastructure. The price of connecting the edge device to AWS Lambda is $p_E$ dollars per month. The cloud resources are following the AWS Lambda resource model in which the user specifies a set of functions $\{f_{i} \mid i = 1 \dots n\}$. For each function the user requests a memory $m_{i,C}$ that will be allocated to the container/VM executing the function $f_i$. The user does not need to worry about the VM executing the function and how they are provisioned, the pricing only depends on the memory and the duration as we will describe in the pricing model (Section~\ref{sec:price}). We note that there are other resources allocated to the function such as timeout, and maximum concurrency. However, in this work we only focus on the memory resource and we assume that the function runs in finite duration and within a maximum concurrency equivalent to the default concurrency (max concurrency = 1000) which means that the function cannot serve more than 1000 requests at a time.  

\subsection{Data model:} The data sent to the Lambda function is in the form of a request encoded in JavaScript Object Notation (JSON) format. Since JSON is a text-based format, it can directly encode text data such as text files or sensor readings. However, if the data is in binary format such as a compressed image,  then the image is first uploaded to a persistent storage (e.g., Amazon S3) and the JSON request will encode the location  of the uploaded image as shown in the example JSON request in Figure~\ref{fig:json}. The number of requests per month is denoted by $r$. We do not make any explicit assumptions on whether the requests come as a continuous stream or they come in bursts.  
\begin{figure}
\begin{lstlisting}[language=json,firstnumber=1]
{
  "userId": "user_1",
  "s3Bucket": "face-collection-bucket",
  "s3Key": "face_photo.jpg"
}
\end{lstlisting}
\caption{Example input request to Lambda function}
  \label{fig:json}
\end{figure}
\subsection{Workflow model:}
We model the workflow in AWS Lambda as a directed-acyclic-graph (DAG) $G_f=(V_f,E_f)$ of functions (Fig.~\ref{fig:task_graph}). A function is a processing element that can execute user-defined code (e.g. convolution, face detection). The vertices in the DAG represent functions $V_{f} = \{f_{i} \mid i = 1 \dots n\}$, and the links between them $E_{f} = \{f_{i} \rightarrow f_{j} \mid i \neq j\ , \ 1 \le i,j\le n\}$ represent the workflow dependencies, where $f_i \rightarrow f_j$ means $f_i$ is executed before $f_j$ and the output of $f_i$ is the input of $f_j$.

\begin{table}[htbp]\caption{Table of Notation}
\begin{center}
\begin{tabular}{r c p{10cm} }
$n$ & $=$ & Total number of functions \\
$r$ & $=$ & Total number of executions of a workflow \\
$G_f$ & $=$ & Input function graph \\
$G'_f$ & $=$ & Fused function graph\\
$f_i$ & $=$ & Function $i$ in graph $G_f$ \\
$f'_i$ & $=$ & Fused function $i$ in graph $G'_f$ \\
$X_i$ & $=$ & Placement variable ( 1: $f_i$ on cloud, 0: $f_i$ on edge)  \\
$t_{i}$ & $=$ & Completion time of function $i$\\
$e_{i,C}$ & $=$ & Execution time of function $i$ on the cloud \\
$e_{i,E}$ & $=$ & Execution time of function $i$ on the edge \\
$D_{f_i}$ & $=$ & Size (bytes) of output data of function $f_i$ \\
$B_{E,C}$ & $=$ & Bandwidth (bytes/sec) between edge and cloud \\
$tr(E\xrightarrow{\text{$D_{f_i}$}}C)$ & $=$ & Transmission time (sec) between edge and cloud \\
$s_{i,C}$ & $=$ & Time to schedule function $i$ on the cloud \\
$m_{i,C}$ & $=$ & Memory allocated to function $i$ \\
$m_{i}$ & $=$ & Maximum memory used by function $i$ \\
$p_E$ & $=$ & Price of connecting one edge device to AWS cloud\\
$p_s$ & $=$ & Price of one state transition\\
$p_{m_{i,C}}$ & $=$ & Price of 1 sec exec. of function $i$ with memory $m_{i,C}$\\
$P(G_f,{X_{i=1,..,n}})$ & $=$ & Price of workflow $G_f$ according to ${X_{i=1,..,n}}$   \\
$T(G_f,{X_{i=1,..,n}})$ & $=$ & Execution time of $G_f$ according to ${X_{i=1,..,n}}$   \\

\end{tabular}
\end{center}
\label{tab:notation}
\end{table}

\subsection{Function Profile}
Each function in Figure~\ref{fig:workflow} is associated with a {\it profile} which includes:
\begin{enumerate}
\item The cost of executing function $f_i$ when placed on node $E$ or $C$. Function $f_i$ can be placed on node $E$ or $C$ so we denote their corresponding execution costs as $e_{i,E}$ and $e_{i,C}$. We assume that each request has equal sized data (e.g., 720p images), therefore the execution cost is the same for all requests.

\item The size of output data of function $f_i$ is denoted as $D_{f_i}$ bytes. The transmission time of the result of $f_i$ from the edge to the cloud is $tr(E\xrightarrow{\text{$D_{f_i}$}}C) = \frac{D_{f_{i}}}{B_{E,C}}$, where $B_{E,C}$ is the bandwidth (bytes/sec)  from the edge to the cloud.

\item The maximum memory consumed by function $f_i$, is denoted as $m_i$. We note that AWS Lambda has only a discrete set of memory values that can be allocated to a function (e.g., 128 MB, 256MB, 320MB, 384MB,...). If the actual memory consumption $m_i$, reported by AWS, is not equivalent to any of the allowed values, a user has to set allocated memory $m_{i,C}$ to the closest allowed value that is larger than $m_i$. Therefore, if $m_i=340MB$, we assume that the allocated memory is $m_{i,C}=384MB$.

\item The scheduling delay $s_{i,C}$ of the function. When AWS receives requests to run a function, it reports the timestamps about the following timed events: (1) Request received, (2) Function scheduled for execution, and (3) Function started execution. The scheduling delay is the time between receiving the request and starting to execute the function.

\end{enumerate}

\subsection{Price Model}\label{sec:price}

For each function $f_i$, we define the variable $X_{i}$ which takes a binary value. $X_{i}$ equals to 1 when $f_i$ is executed on the cloud and it takes a 0 value when $f_i$ is executed on the edge device.
We note that the edge device has a fixed cost $p_E$ for connecting it to the AWS cloud no matter how many functions are allocated to it. On the other hand, the price of executing $f_i$ on the cloud depends on the memory $m_{i,C}$ allocated to it, and its execution time $e_{i,C}$. The price per 1 GB memory and 1 sec of execution time is denoted as $p_{m_{i,C}}$. The price for each state transition (i.e., each link in Figure~\ref{fig:workflow}) is denoted as $p_s$. We formulate the price per month $P$ as follows:

$$P(G_f,{X_{i=1,..,n}})=\sum_{i=1}^{i=n} X_{i} \cdot r \cdot e_{i,C} \cdot m_{i,C} \cdot p_{m_{i,C}} + r \cdot (n+1) \cdot p_s + p_E $$

We note than the number of transitions for $n$ functions is $n+1$ by calculating the start and end transitions as shown in Figure~\ref{fig:workflow}, entire workflow of functions is executed for $r$ times then the number of transitions is also multiplied by $r$, so the total number number of executed transitions is $r(n+1)$.
We also note that AWS offer some requests and transitions free of charge in the beginning of each month but our pricing model only considers the price after the free requests are consumed.


\subsection{Execution Time Model}\label{sec:exec-time}
In this section, we formulate the execution time for each request.  The execution time for a request is defined by the completion time of the last function $f_n$ minus the starting time of the first function $f_1$. We denote the completion time of $f_n$ as $t_{n}$. The total execution time $T$ of the workflow is given by:

$$T(G_f,{X_{i=1,..,n}})=t_n(G_f,{X_{i=1,..,n}}) - t_0$$

such that $t_0$ is the time before starting the execution of $f_1$. The completion time of function $f_i$ is given by the recursive formula:

\begin{equation}
\begin{split}
t_i(G_f,{X_{i=1,..,n}}))=\underbrace{t_{i-1}(G_f,{X_{i=1,..,n}})}_\textrm{completion time of prev. function} + \underbrace{(1 - X_{i}) \cdot e_{i,E}}_\textrm{Execution time on edge} \\ + \underbrace{\mid X_{i} - X_{i-1} \mid \cdot tr(E\xrightarrow{\text{$D_{f_{i-1}}$}}C)}_\textrm{Transmission time}  + \underbrace{X_i \cdot (e_{i,C} + s_{i,C})}_\textrm{Execution time on cloud}
\end{split}
\end{equation}

Such that $tr(E\xrightarrow{\text{$D_{f_{i-1}}$}}C)$ is the transmission time of the intermediate data from $f_{i-1}$ to $f_i$. We assume that the intermediate data between two functions running on the cloud or two functions running on the edge is negligible. 

\subsection{Problem Definition:} Let $G_f'=(V_f',E_f')$ be the new function graph after function fusion, such that  
the vertices represent fused functions $V'_{f} = \{f_{i}' \mid i = 1 \dots m\}$ and each fused function is a concatenation of two or more functions $f_{i}'=f_1 \mid f_2 \mid f3 \mid ..$, where the symbol "$\mid$" denotes concatenation. Let $X_i'$ be the placement variable $\{X_i' \mid i = 1 \dots m\}$ for each fused function $f_i'$. Given the function graph $G_f$, we define the cost optimization problem as finding the fused graph $G_f'$ and the placement variables $\{X_i' \mid i = 1 \dots n\}$ such that the price $P_{G_f', {X_{i'=1,..,m}}}$ is minimized and the execution time does not exceed a certain threshold $T_{thresh}$ so the problem can be formulated  as:

$\displaystyle{\minimize\ P_{G_f', {X_{i'=1,..,m}}}}$
where
$T_{G_f', {X_{i'=1,..,m}}} < T_{thresh}$

%% file: approach.tex
\section{Proposed Approach}\label{sec:approach}
\begin{figure}
\centering
  \includegraphics[width=0.3\textwidth]{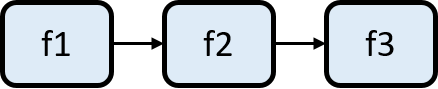}
  \captionof{figure}{Example Workflow with three functions }
  \label{fig:task_graph}
\end{figure}

One of the key contributions in our approach is to jointly represent the solutions for the function placement and function fusion in one graph which we refer to as the \emph{Cost graph}. In order to illustrate how we build the \emph{Cost graph}, we take an example function graph $G_f$ with three functions$V_f={f_1,f_2,f_3}$ as shown in Figure~\ref{fig:task_graph} and two different resources: one edge device $E$ and one cloud node $C$. We assume that intermediate output data can flow from $E$ to $C$ but not vice versa. 

\begin{figure}
 \centering
  \includegraphics[width=0.2\textwidth]{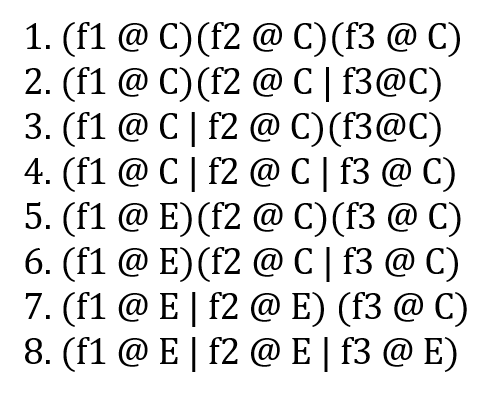}
  \captionof{figure}{Feasible function placement and function fusion solutions. Each line is one solution. Symbol "$\mid$" denotes fusion }
  \label{fig:solutions}
\end{figure}

{\bf Feasible Solutions:} Each solution requires deciding which functions to fuse if any (Function fusion) and assigning each fused function to $E$ or $C$ (Function placement). We show the possibilities of function fusion and placement in Figure~\ref{fig:solutions}. Each row is a possible solution to the cost optimization problem. The label $f_1 @ E$ means that $f_1$ is placed on device $E$, similarily the label $f_1 @ C$ means that $f_1$ is placed on the cloud. The functions inside parentheses show the functions that are being fused and the symbol "$\mid$" denotes fusion operation. 

Line 1 in Figure~\ref{fig:solutions} shows the possibility that the three functions remain unchanged and placed on the cloud. Lines 2-4 show different possibilities of fusing functions while remaining on the cloud. Lines 5-7 show some possibilities for partitioning the functions across $E$ and $C$. Line 8 shows an extreme case when all functions are fused together and placed on the edge. We note that we prune some of the solutions such as $(f_1 @ E)(f_2 @ E)(f_3 @ E)$ because it is equivalent to line 8 (i.e., fusing the three functions and placing them on the edge). We note that placing three functions on $E$ without fusing them neither have a price or execution time benefit. We also prune solutions like $(f_1 @ C)(f_2 @ E \mid f_3 @ E)$ because the data flows from $E$ to $C$ but not vice versa.

{\bf Cost Graph Representation:} Instead of enumerating the possible solutions in Figure~\ref{fig:solutions}, we construct a {\it cost graph} as shown in Figure~\ref{fig:costgraph}. Each node represents one unique parenthesis from Figure~\ref{fig:solutions} and is labeled by a fused function. A link from node $u$ to node $v$ in the cost graph corresponds to the ending of one pair of parenthesis and the beginning of the next one in Figure~\ref{fig:solutions}. A link in the cost graph represent the transition from one fused function to another and it holds two independent costs: 

\begin{figure}
  \includegraphics[width=0.48\textwidth]{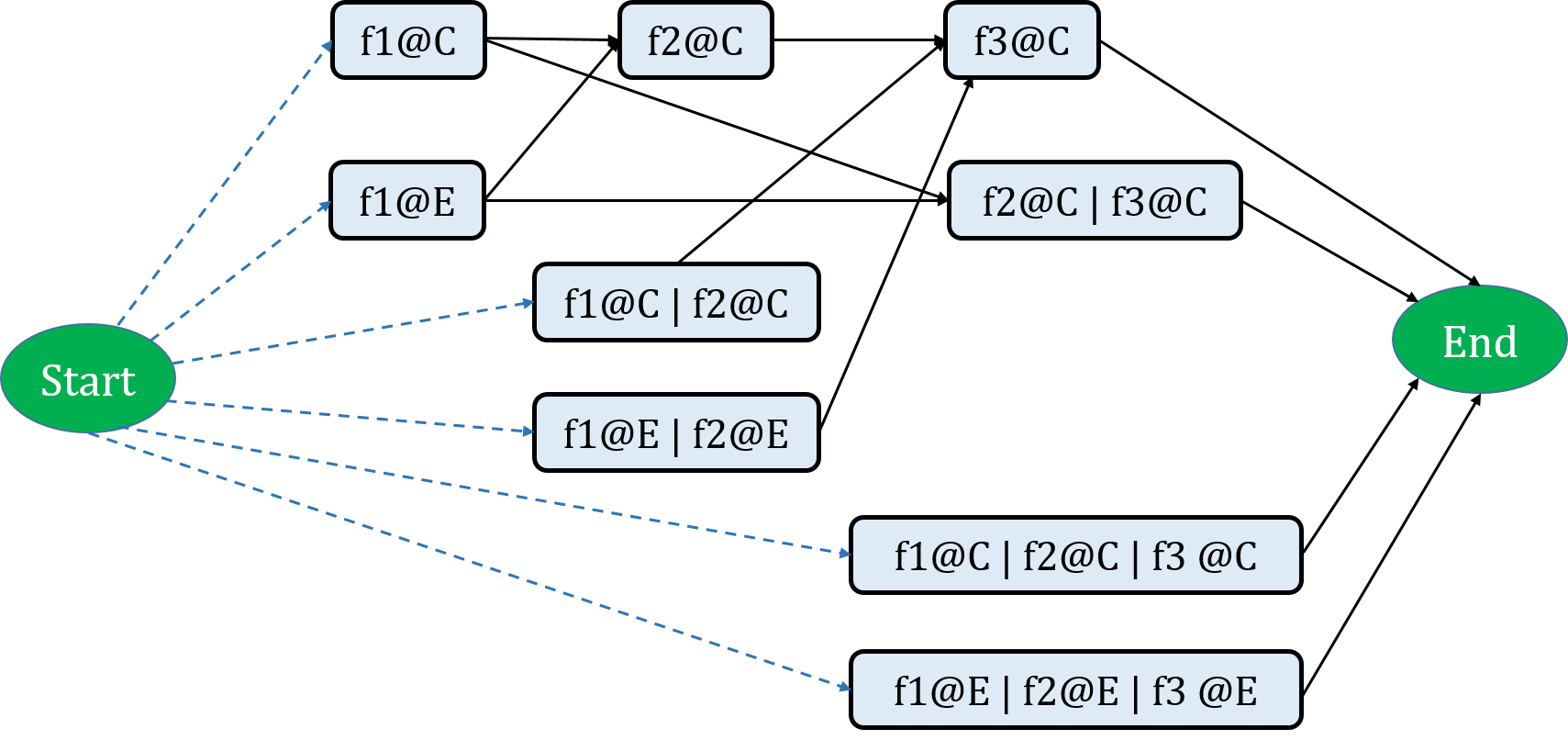}
  \captionof{figure}{ Cost Graph, each path represent a solution for function placement and fusion. Each edge include execution time and price costs }
  \label{fig:costgraph}
\end{figure}

\begin{enumerate}

\item {\bf Price Cost $c_{uv}$}: this includes the price of fused function $i$ and the state transition from $i$ to $j$.

\item {\bf Delay Cost $d_{uv}$}: the execution time of fused function $i$ and the transmission time of output data from $i$ to $j$. 

\end{enumerate}

\begin{figure*}
 \centering
  \includegraphics[width=0.7\textwidth]{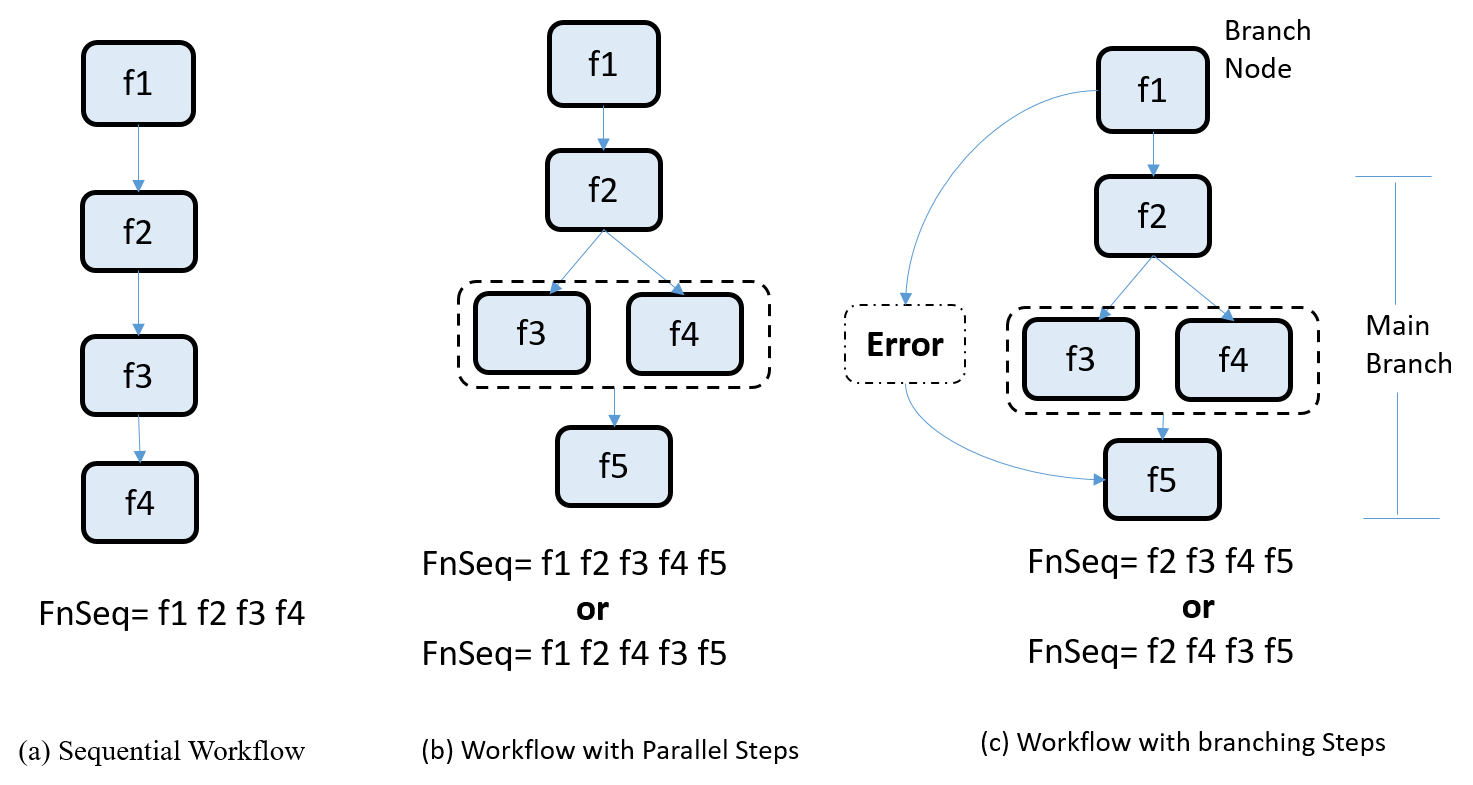}
  \captionof{figure}{Different workflow types supported by AWS }
  \label{fig:different_workflows}
\end{figure*}

We will discuss more details about constructing the placement graph and calculating the link weights in Sections~\ref{sec:steps} and ~\ref{sec:cost} .

Intiuitively, the shortest path between the start and end nodes is the solution of the cost optimization problem. For example, if the shortest path is $[(f_{1} @ E)(f_2 @ C \mid f_3 @ C)]$, then $f_{1}$ is placed on device $E$ while $f_{2}$ and $f_{3}$ are fused into one function and placed on the cloud. The challenge to solve this problem is that each link has two independent costs (i.e., price and time) which makes it infeasible to solve the problem using the standard shortest path algorithms such as Dijkstra~\cite{DjikstrasAlgo}. Therefore, we formulate the problem as a constrained shortest path problem (CSP).




{\bf Problem Transformation (Constrained Shortest Path):}

Let us consider a cost graph $G_c(V_c,E_c)$, each link $(u,v) \in E_c$ is associated with two costs: a price cost $c_{uv}$, and a delay cost $d_{uv}$. Let $s$ and $t$ be two distinguished nodes in the graph. For each path, let us denote the set of all paths from $s$ to $t$ as $Y_{st}$. For any path $y \in Y_{st}$, we define:
$$c(y)=\sum\limits_{(u,v) \in y} c_{uv} \ and \ d(y)=\sum\limits_{(u,v) \in y} d_{uv}$$

Given $T_{thresh} > 0$, let $\overline{Y}_{st}$ be the set of all s-t paths y such that  $d(y) \leq T_{thresh}$. The CSP problem is to find a path $y^*=arg min\{c(y) \mid y \in \overline{Y}_{st}\}$. In other words the problem is to find the path with the minimum price such that the delay does not exceed a threshold $T_{thresh}$.


\subsection{Costless Algorithm Steps:}\label{sec:steps}
In this section, we described the steps of Costless which include handling different workflow types, constructing the cost graph, and solving the constrained shortest path problem.

{\bf Step 1: Create  an intermediate representation from different workflow types:} AWS Lambda provides an API to define the the application workflow as a series of steps. Figure~\ref{fig:different_workflows} shows examples of the flow of steps which include sequential, parallel, and branching steps. We try to represent the workflow as a sequence of functions $FnSeq$. 
Figure~\ref{fig:different_workflows}(a) shows the simplest case which is the sequential steps. We can represent the $FnSeq$ based on the order of functions from $Start$ to $End$. Figure~\ref{fig:different_workflows}(b) shows a workflow with parallel steps in which multiple functions are executed in parallel and their outputs are aggregated before the next function starts. Similar to the sequential workflow, we represent the $FnSeq$ based on the order of functions from start to end and we arbitrarily order the parallel steps. We note that in addition to the $FnSeq$, we keep track of the original DAG representation while calculating the cost in the cost graph to take the maximum execution time as the reference execution time for the entire parallel step, and we consider the sum of the prices of parallel function as the price for the entire parallel step. 

The last type of workflows that we deal is the workflow with branching steps shown in  Figure~\ref{fig:different_workflows}(c). Such workflows typically have a branch node that has a condition to decide in which branch the execution will proceed. Since, we are interested in optimizing the cost for the general case, we only focus on the main branch that contains the functions rather than the branch that contains the error handling. Therefore, we represent the $FnSeq$ based on the main branch only. In Figure~\ref{fig:different_workflows}(c), the main branch is similar to the workflow with parallel steps in Figure~\ref{fig:different_workflows}(b) and the $FnSeq$ representation follows the same method. 


{\bf Step 2: Construct cost graph:} The list $FnSeq$ obtained in Step 1 has a sequence of functions that follow certain execution order. We use parenthesis around adjacent functions in $FnSeq$ to create fused functions of size up to the length of $FnSeq$ For example, the set of possible fused functions $F'$ for $FnSeq=\langle f_{1},f_{2},f_{3}\rangle$ are:

$F'=\{\underbrace{(f_1) , (f_2), (f_3)}_\textrm{orginal non-fused functions}, \underbrace{(f_1\ f_2) , (f_2\ f_3)}_\textrm{fusing two functions},\underbrace{(f_1\ f_2\ f_3)}_\textrm{fusing three functions}\}$

We define a function $L(f')$ that denotes the set of possible placements of function $f'$. In general, each function $f'$ that starts with $f_1 $ can be placed on either edge device $E$ or cloud $C$. However, if $f'$  does not start with $f_1$ then it can only be placed on $C$ because the data does not flow from $C$ to $E$. In the following we show some examples of $L(f')$
$$L(f'=(f_1)) = \{ f_1 @ E, f_1 @ C \}$$
$$L(f'=(f_2\ f_3)) = \{({f_2} @ C \mid {f_3}@C)\}$$
We use the placement sets to generate a cost graph $G_c = (V_c,E_c)$ (See Figure~\ref{fig:costgraph}), where the vertices are the union of all placement sets: $V_c={\cup_{f'} L(f')}$. The following are the values of the rest of the placement sets for the three functions example:
\begin{align}
&\begin{aligned}
L(f'=(f_2)) = \{ f_2 @ C \}  
\end{aligned}\\
&\begin{aligned}
L(f'=(f_3)) = \{ f_3 @ C \}
\end{aligned}\\
&\begin{aligned}
L(f'=(f_1\ f_2)) = \{ ({f_1} @ E \mid {f_2}@E), ({f_1} @ C \mid {f_2}@C)\}
\end{aligned} \\
&\begin{aligned}
L(f'=(f_1\ f_2\ f_3)) = \{ ({f_1} @ E \mid {f_2}@E \mid {f_3} @ E), \\({f_1} @ C \mid {f_2}@C \mid {f_3} @ C) \}
\end{aligned}
\end{align}
%
%

We note that in order to try two memory configurations $m_1$ and $m_2$ on the cloud the placement set will have more combinations for example: 

\begin{equation*}
\begin{split}
L(f'=(f_1\ f_2)) =  \{({f_1} @ E \mid {f_2}@E), ({f_1}  @ C_{m_1} \mid   {f_2}@C_{m_1}), \\ ({f_1} @ C_{m_2} \mid {f_2}@C_{m_2})\}
\end{split}
\end{equation*}



{\bf Step 3: Add cost graph links}
Once we have the nodes, as shown in Figure~\ref{fig:costgraph}, we start by adding links from the $START$ node to all nodes that start with first function in $FnSeq$, e.g. $f_{1}$. Similarly, we add links for all nodes that end with last function in $FnSeq$ (i.e., $f_{3}$), to the $END$ node. To add links between intermediate nodes, we add links between each node that starts with function $i$, and each child node that starts with function $j$, where $j$ is the successor of $i$ in $FnSeq$. We discuss the details of calculating link weights in Section~\ref{sec:cost} 

{\bf Step 4: Solve the CSP problem:} The constrained shortest path (CSP) problem is known to be NP-hard~\cite{csp}. However, several approximation algorithms have been proposed to solve it~\cite{csp}\cite{csp1}\cite{csp2}. Out of these methods, the LARAC algorithm~\cite{csp} which is based on a relaxation of the CSP problem is an efficient algorithm to solve the problem. The main idea behind LARAC is to apply Dijkstra's shortest path algorithm on an aggregated cost $c_{uv}/c^*+\lambda d_{uv}/d^*$ that includes both the price and the delay values. The key  
issue in solving the CSP problem becomes how to search for the optimal 
$\lambda$ and determining the termination condition for the search. LARAC provides an efficient search procedure. We note that $c_{uv}$ and $d_{uv}$ are measured in different units (\$ and seconds). Therefore, $c_{uv}$ and $d_{uv}$  are normalized through dividing them by $c^*$ and  $d^*$ which are the maximum cost and delay values.


\subsection{Cost Calculation:}\label{sec:cost}
In this section, we show how we calculate the cost of the links in the cost graph. We take an example path: $$(f_{1} @ E) \rightarrow (f_{2} @ C \mid f_3 @ C) \rightarrow (f_4 @ C) \rightarrow End$$ The path consists of 3 functions and the middle function is a fused function that consists of $f2$ and $f3$. We calculate both price and execution time cost for the link between each two consecutive functions. The price cost of the links on the path are added together to form the price cost of entire path. Similarly, the execution time of a path is the sum of executions times of the links on the path. In the following we focus on calculating the price $P$ and the execution time $T$ for each link and we follow the notation in Table~\ref{tab:notation}
\begin{enumerate}
\item {\bf Cost of link $(f_{1} @ E)\rightarrow(f_{2} @ C \mid f_3 @ C)$}:
$$T[(f_{1} @ E)\rightarrow(f_{2} @ C \mid f_3 @ C)] = \underbrace{e_{1,E}}_\textrm{execution cost} + \underbrace{ tr(E\xrightarrow{D_{f_{1}}} C) }_\textrm{transmission cost}$$

$$P[(f_{1} @ E)\rightarrow(f_{2} @ C \mid f_3 @ C)] = \underbrace{ p_E}_\textrm{edge device price } + \underbrace{ r \cdot p_s}_\textrm{transition price}$$

\item {\bf Cost of link $(f_{2} @ C \mid f_3 @ C) \rightarrow (f_{4} @ C)$}:
We note that the cost of this link is different based on whether $f_2$ and $f_3$ are parallel functions or not, if they are parallel functions then the execution time will be bounded by the slowest functions, otherwise the execution time of the fused function will be the sum of execution time $f_2$ and $f_3$. The following is the calculation for both cases:

{\bf If $f_2$ and $f_3$ are parallel:}

$$T[(f_{2} @ C \mid f_3 @ C) \rightarrow (f_{4} @ C)] = \underbrace{max(s_{2,C} + e_{2,C},s_{3,C} + e_{3,C})}_\textrm{scheduling and execution time}$$ 
\\

$P[(f_{2} @ C \mid f_3 @ C) \rightarrow (f_{4} @ C)] = \underbrace{ r \cdot (e_{2,C} \cdot m_{2,C} \cdot p_{m_{2,C}})  + r \cdot (e_{3,C} \cdot m_{3,C} \cdot p_{m_{3,C}} )}_\textrm{functions price}  + \underbrace{ 2 \cdot r \cdot p_s}_\textrm{transition price}$
\\
\\
{\bf if $f_2$ and $f_3$ are NOT parallel:}
\\
$$T[(f_{2} @ C \mid f_3 @ C) \rightarrow (f_{4} @ C)] = \underbrace{s_{2,C} + e_{2,C} + e_{3,C}}_\textrm{scheduling and execution time}$$

$P[(f_{2} @ C \mid f_3 @ C) \rightarrow (f_{4} @ C)] = \underbrace{ r \cdot (e_{2,C} + e_{3,C}) \cdot max(m_{2,C},m_{3,C}) \cdot p_{max(m_{2,C},m_{3,C})}}_\textrm{fused function price}  + \underbrace{r \cdot p_s}_\textrm{transition price}$
\\

When $f_2$ and $f_3$ are not parallel, then they can be fused together which implies that: (1) they incur only one scheduling delay; (2) their execution times are added to each other; (3) the memory of the fused function is the maximum of the memory allocated to individual functions, and (4) they have one output transition.

\item {\bf Cost of link $(f_{4} @ C) \rightarrow END$}:

$$T[(f_{4} @ C) \rightarrow END] = \underbrace{s_{4,C} + e_{4,C}}_\textrm{scheduling and execution time}$$

$$P[(f_{4} @ C) \rightarrow END] = \underbrace{ r \cdot (e_{4,C} \cdot m_{4,C} \cdot p_{m_{4,C}})}_\textrm{functions price}  + \underbrace{ \cdot r \cdot p_s}_\textrm{transition price}$$

\end{enumerate}

\subsection{Algorithm analysis}
Among the 4 steps described in section~\ref{sec:steps}, solving the constrained shortest path (CSP) problem (Step 4) is the one that dominates the algorithm complexity.
The complexity of LARAC's algorithm~\cite{larac-complexity} for solving CSP on the cost graph is given by: 

\begin{equation}\label{eq:larac}
O(\mid E^2 \mid log^2(\mid E \mid))
\end{equation}
such that $\mid E \mid$ is the number of links in the cost graph which is defined as.

\begin{equation}\label{eq:links}
\mid E \mid = \mid V \mid \cdot deg_v 
\end{equation}

such that $\mid V \mid$ is number of vertices in the cost graph, $deg_v$ is the maximum out degree of each vertex. In order to calculate these values, we denote $m$ as the number of devices in which the function can be placed on, where $m=2$ for one edge device and one cloud configuration, and $m=4$ for one edge and 3 cloud configurations. $\mid V_p \mid$ is calculated as the number of fused functions multiplied by the number of possible placements (i.e., $m$) of each fused function (See Step 2 in section~\ref{sec:steps}). Given $n$ functions, the number of fused functions of size $k$ is given by $(n-k+1)$, therefore the total number of vertices $\mid V \mid$ in the cost graph can be written as:

\begin{equation}\label{eq:vertices}
\mid V \mid = \sum_{k=1}^{n} (n - k + 1) \cdot m
\end{equation}

The worst case out degree of cost graph nodes is:

\begin{equation}\label{eq:deg}
deg_v=(n-1)\cdot m
\end{equation}

This happens for nodes $f1@C$ and $f1@E$ which are connected to $(n-1)$ fused functions that starts with $f2$ and for each fused function there are $m$ nodes that represent each possible placement of the fused function. By substituting Equations~\ref{eq:vertices} and~\ref{eq:deg} in Equation~\ref{eq:links}, the number of links can be defined as:

\begin{equation}\label{eq:links-detail}
\begin{split}
\mid E \mid = m^2 \cdot (n-1) \cdot \sum_{k=1}^{n} (n - k + 1) =  m^2 \cdot (n-1)  (\sum_{k=1}^{n} n - \sum_{k=1}^{n} k \\   + \sum_{k=1}^{n} 1) =   m^2 \cdot (n-1) (n^2 - (n^2 + n)/2 + n) = O(m^2 \cdot n^3)
\end{split}
\end{equation}

By substituting Equation~\ref{eq:links-detail} in~\ref{eq:larac}, the overall complexity is given by:

\begin{equation}\label{eq:overall}
\begin{split}
O(\mid E^2 \mid log^2(\mid E \mid)) = O(m^4 \cdot n^6 \cdot log^2(m^2 \cdot n^3))
\end{split}
\end{equation}

%% file: eval.tex
\section{Evaluation}\label{sec:eval}

{\bf Experimental Setup:} The computing infrastructure consists of edge and cloud resources. We use the Raspberry Pi Model B as the edge device. For the cloud side we use AWS Lambda and AWS Step Functions to create a workflow of lambda functions. We set the default memory of each lambda function according to the closest values to the maximum memory used by the function. For example, if the function profiling shows that a function uses a maximum of 100 MB of memory, we allocate memory to be the closest allowed value by AWS Lambda which is 128MB. We set the timeout to a large value to keep the function running until it finishes execution. We assume that the data comes from the Raspberry Pi and it is uploaded to AWS Storage (S3) for cloud processing. Once the data is uploaded to S4, it automatically triggers the execution on the cloud. If a function is placed in the edge, it is executed first and the intermediate data is transmitted to the cloud.

\begin{figure}
  \includegraphics[width=0.42\textwidth]{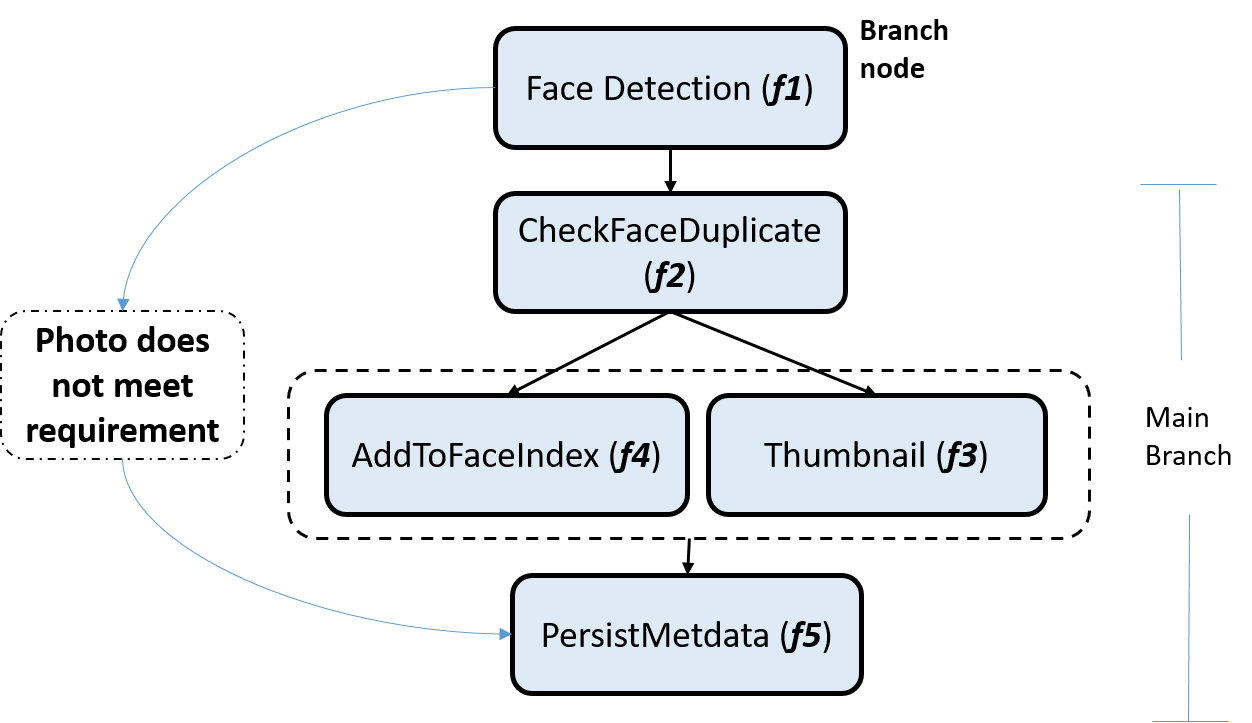}
  \captionof{figure}{Wild Rydes application workflow}
  \label{fig:wild}
\end{figure}

{\bf Application:} We evaluate the performance of our algorithm using Wild Rydes application workflow~\cite{wild}. 
Wild Rydes is a transportation application similar to Uber that allows users to request rides in an on-demand manner and the application matches them with the closest drivers. Wild Rydes requests its users to upload their photo when they sign up for a new account. Once the user uploads their photo, the image processing workflow in Figure~\ref{fig:wild} starts executing. The workflow consists of five functions that process the image, matching it across a database of faces  and indexing the uploaded face for future matching. The workflow is implemented in $JavaScript$ and it takes an image as input. For the sake of brevity, we label the functions from $f1$ to $f5$ as shown in Figure~\ref{fig:wild} and we use the labels for the rest of the paper. $f1$ is a branch function that decides the execution of the rest of the application. As described in section~\ref{sec:steps}, we do not include the branch function $f1$ in the fusion and we explore fusion for the rest of the 4 functions. $f1$ is a face detection function that uses AWS Rekognition library which is a cloud-based library offered by Amazon that contains a variety of image and video processing functions. Since AWS Rekognition does not offer a distribution that can be deployed on a Raspberry Pi, we implemented the face detection functionality using Python's Dlib~\cite{dlib} library. The rest of the functions is implemented on the cloud because the functions need to access two databases of faces and metadata that are stored on the cloud. 


{\bf Application Profiling:} We profile the application by executing the workflow for 20 times on both the edge and the cloud. On the edge, we run the $FaceDetection$ only because it is only function that does not depend on the cloud database. On the cloud we run each function with two memory configurations $m_1=128$ and $m_2=256$. We use AWS logs to extract the following profiling information {\bf for each function}:
\begin{enumerate}
\item Average Execution time on the cloud using 128 MB (Default configuration unless otherwise stated)
\item Average execution time on the cloud using 256 MB
\item Average execution time on the edge
\item Average scheduling delay on the cloud
\item Maximum memory used
\item Average billed duration using 128 MB 
\item Average billed duration using 256 MB
\end{enumerate}

Table~\ref{tab:profile} shows the results of the profiling. We note that the average billed duration is always greater than the average execution time because it is rounded up to the nearest 100 milliseconds. For example if the execution is 720ms, AWS charges for 800ms so the average billed duration tends to be bigger than the average execution. We also notice that the first run is usually much slower than other runs. This is probably due to the fact that AWS provisions a container in the first run that can be reused for other requests. We consider the first run as a warm up run and we do not include it in the profiling.

We further run some benchmarking to measure the transmission time from the Raspberry Pi to the Amazon S3 and we measure the duration from sending the 100 images to receiving a response from S3 that the data has been uploaded successfully. We use this information to measure the speed of transfer from the edge to the cloud and on average it takes $1.13$ seconds to upload a 720p image with sizes between 1.2-1.5 MB size. We note that uploading the image is also needed when $f1$ is executed on the edge because the image is needed by other downstream tasks. 

\begin{table*}[h]
\begin{center}
\begin{tabular}{|c|c|c|c|c|}
\hline \bf Function & \bf Avg. exec. time [128 MB / 256 MB / Edge] & \bf Avg. scheduling delay & Max Memory used & Avg. billed duration [128 MB / 256 MB]    \\ \hline
f1 & 893 ms / 772 ms / 1870 ms & 61 ms & 42 MB & 955 ms / 822 ms \\
\hline
f2 & 970 ms / 743 ms  & 52 ms & 38 MB & 1016 ms / 800 ms \\
\hline
f3 & 2063 ms / 1080 ms  & 172 ms & 83 MB & 2116 ms/1144 ms \\
\hline
f4 & 844 ms / 735 ms  & 153 ms & 37 MB & 883 ms/788 ms \\
\hline
f5 & 153 ms / 101 ms & 67 ms & 38 MB & 211 ms/144 ms \\
\hline

\end{tabular}
\end{center}
\caption{ Profiling information for the functions in the Wild Rydes application (Figure~\ref{fig:wild}). }
\label{tab:profile}
\end{table*}



{\bf Evaluation Metrics:} We conduct experiments to evaluate Costless performance in terms of: 
\begin{itemize}
\item{\it Model Accuracy:} For each fusion and placement solution, we compare our estimate of price and execution time to the observed completion time when we manually fuse and place functions, and we evaluate accuracy of our estimates compared to the billing information from AWS.
\vspace{-2 mm}
\item {\it Price within latency constraint}: Given a function graph and some deadline, we use Costless to find the best 
price for the deadline. We compare the results of Costless with Brute force solution and other heuristics. 

\item {\it Effect of optimizing over memory configurations}: For each fusion and placement solution, we show the price optimization when we search over different memory configurations for each function. 

\item {\it Time to find Solution}: We measure the time that Costless takes to obtain the placement and fusion decision and we focus on its behavior for large scale function graphs. 

\end{itemize}

\begin{figure}
\centering
\includegraphics[width=0.5\textwidth,height=0.35\textwidth]{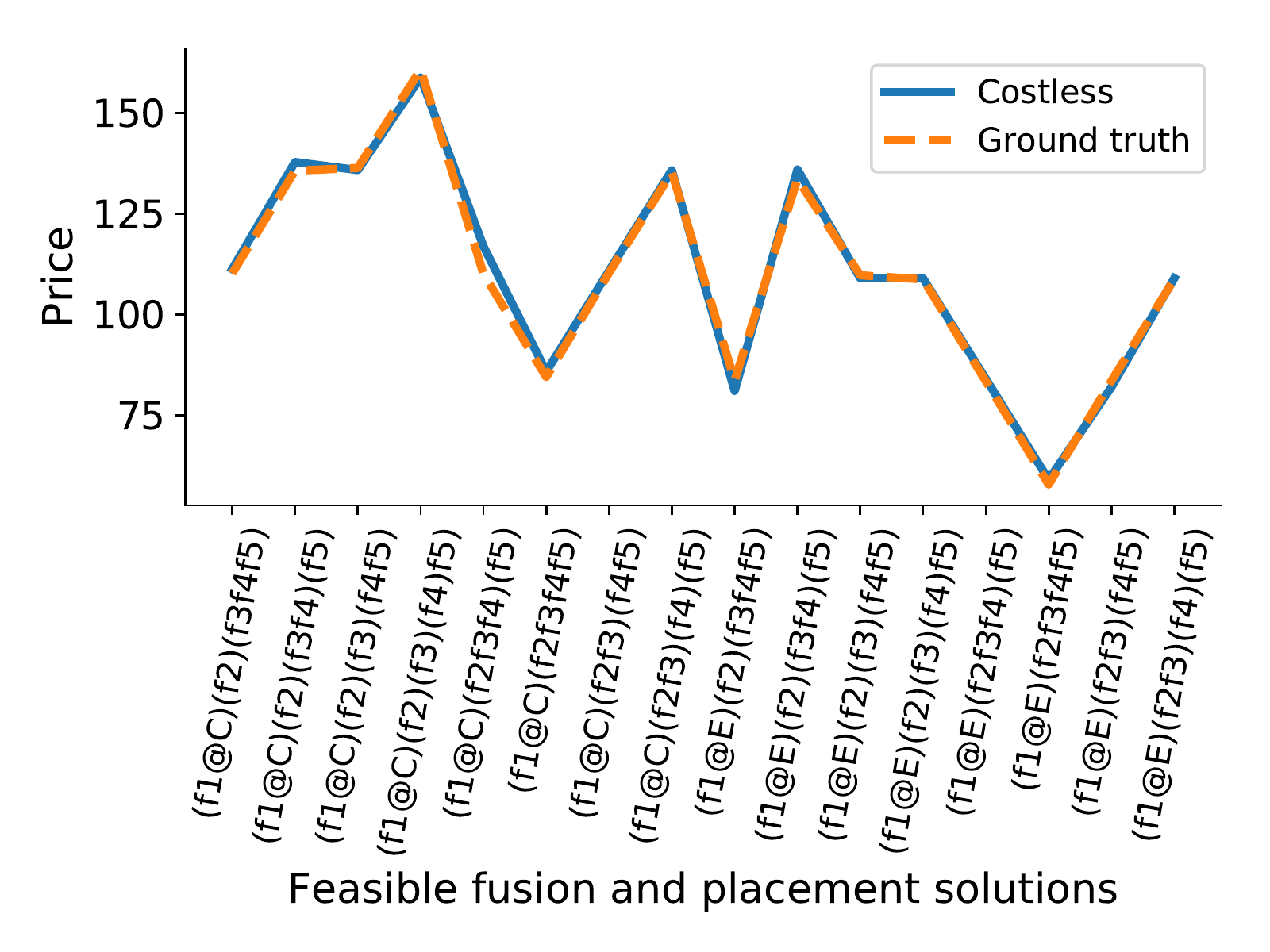}
\caption{Comparing pricing estimates of Costless with observed times during deployment of manually fused functions}
\label{fig:price-model}
\end{figure}

\subsection{Model Accuracy}

\begin{figure}
\centering
\includegraphics[width=0.5\textwidth,height=0.35\textwidth]{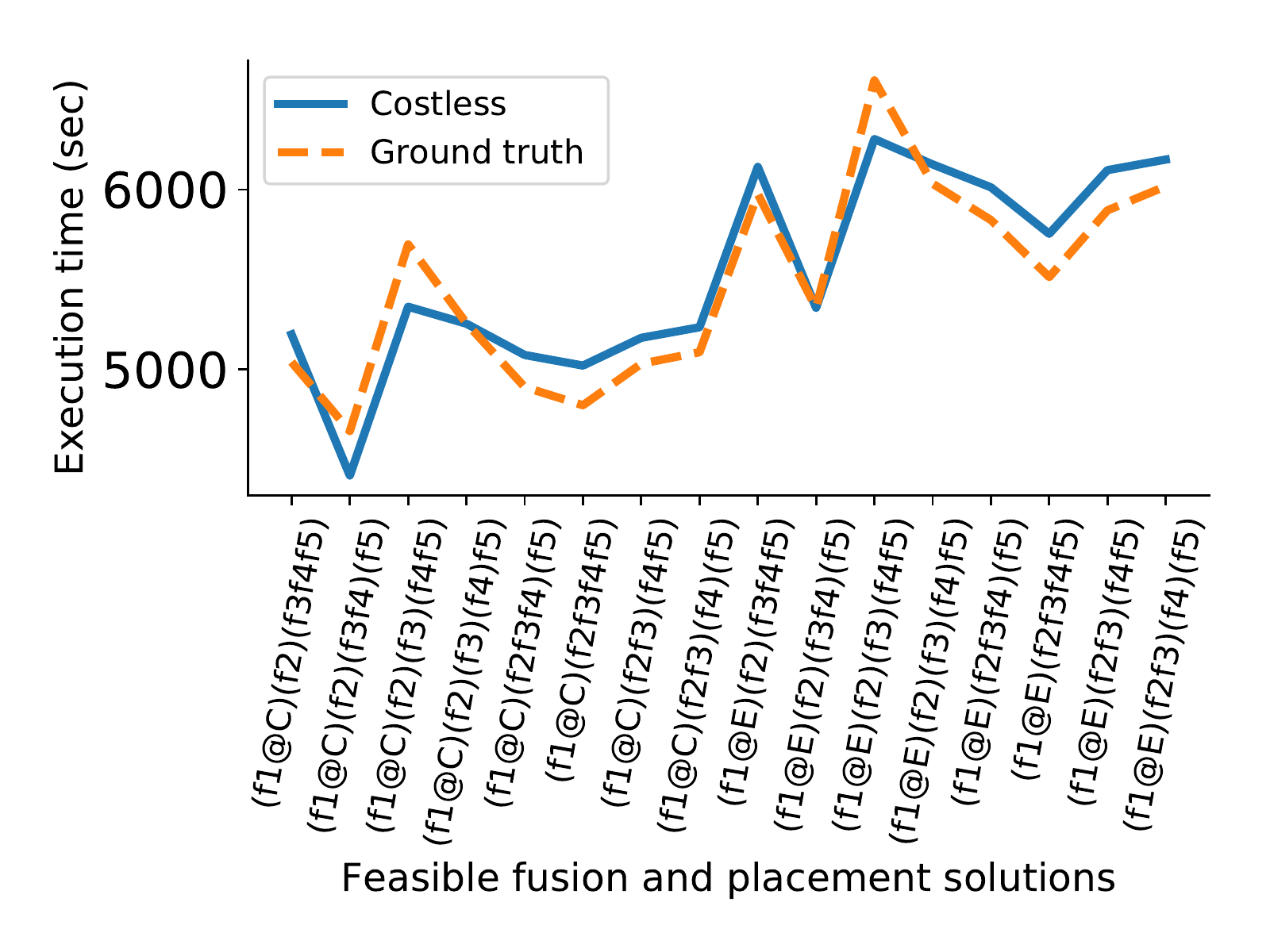}
\caption{Comparing execution time estimate of Costless with observed times during deployment of manually fused functions}
\label{fig:time-model}
\end{figure}

In order to show that our modeling of price and execution closely reflects the observed running time from manually fused functions, we create all feasible function fusions manually for the application in Figure~\ref{fig:wild}. For the 4 functions in the workflow, we try all the combinations of placements and fusions (e.g., fuse two, three or four consecutive functions). We run the manually fused application on AWS and we use logs to find the actual price and execution time and we use this as our {\it Ground truth}. We note that some fusions made the two parallel functions $AddToFaceIndex$ and $Thumbnail$ run sequentially.  Figures~\ref{fig:price-model} and~\ref{fig:time-model} show the results for  the execution time and price estimates. Figure~\ref{fig:price-model} shows that the estimated price and execution for each of the possible solution were very close to the empirically measured values. The execution times in Figure~\ref{fig:time-model} shows slight discrepancies compared to the ground truth obtained from AWS logs, this is due to the fact that there is a scheduling delay between the time of receiving the request and the time at which the function starts execution.
Such delay is highly variant from one request to another and it causes the execution time to deviate by 100-300ms. The execution time estimate, however, follows the same trend as the ground truth and it could capture all the peaks. The average error was only 1.2\% for price and 4\% for execution time.

\subsection{Price vs. Execution time}
\begin{figure}
\centering
\includegraphics[width=0.5\textwidth,height=0.35\textwidth]{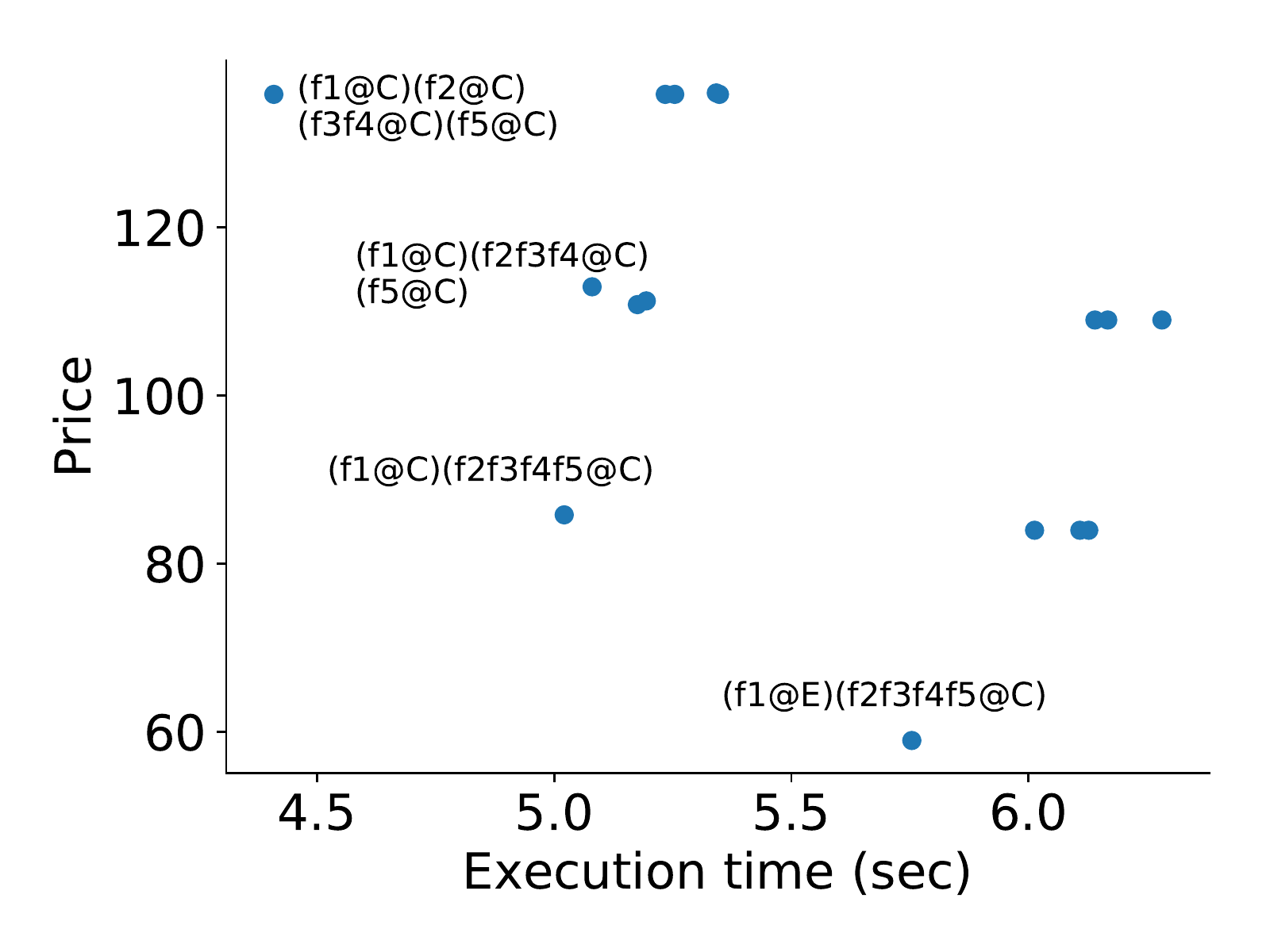}
\caption{Time and Price estimate for each feasible fusion and placement solution obtained by Costless }
\label{fig:price_vs_time}
\end{figure}

In this experiment, we show the relationship between the price and execution time of feasible fusion and placement solutions obtained by Costless. Figure~\ref{fig:price_vs_time} depicts the price execution time relationship. As shown in Figure~\ref{fig:price_vs_time}, there is no clear trend that when the price increases the execution time decreases. In fact, different fusions can give different prices and for each price point there can be different execution times. This due to the fact that some fusions can cause two parallel functions to run sequentially so the price decreases but the execution time increases. In Figure~\ref{fig:price_vs_time}, we focus on four data points that are the most interesting because they have the smallest execution times but yet they have different prices. We note that the most expensive is the original graph in Figure~\ref{fig:wild} and thats because it has four transitions apart from the start and end transition that are common in all solutions. Though it is the most expensive, it has the best execution time because it can leverage the parallelism between the two functions $f3$ and $f4$. On the other hand the cheapest solution is to place $f1$ on the edge and fuse the functions $f2f3f4f5$, fusing the functions not only decreases the price but it also eliminates the need for a scheduling delay between consecutive functions because now several functions became just one function with one scheduling interval so the application runs faster. There are two points in between the cheapest and the most expensive, one of them is very similar to the cheapest solution but it does not use edge device and the other one is more expensive because it fuses 3 functions instead of 4. We note these intermediate solutions that Costless quantifies are helpful because they tend to fuse less so the application retains some of its modularity while improving the price.

\subsection{Price within latency constraint}
In this section we show the benefits of Costless to optimize the price within latency constraint. The goal is that for some deadline (e.g., 5 seconds), we need to find the solution with the lowest price. We compare Costless with several heuristics to assess its ability to accurately find the lowest price solution and we also show it improves on simple decisions that supports using cloud only or edge only. The heuristics compared are:
\begin{itemize}
\item Ground truth: these are the results obtained from manually fusing the functions and running them on AWS. The logs of AWS are automatically parsed to find out the lowest price for a given deadline.

\item Costless: this is the proposed approach in this paper and it uses the price and execution time estimates described in section~\ref{sec:models} and the algorithm described in section~\ref{sec:steps}

\item Bruteforce: this approach uses the price and execution estimate proposed by Costless but instead of constructing a cost graph and using the constrained shortest path approximate algorithm, it searches over all the solutions in a brute force fashion.

\item Cloud (No Fusion): Keep the original application with no fusions and place all the functions in the cloud  

\item Edge (No Fusion): Keep the original application with no fusions and place functions on the edge device whenever possible

\end{itemize}

\begin{figure}
\centering
\includegraphics[width=0.5\textwidth,height=0.35\textwidth]{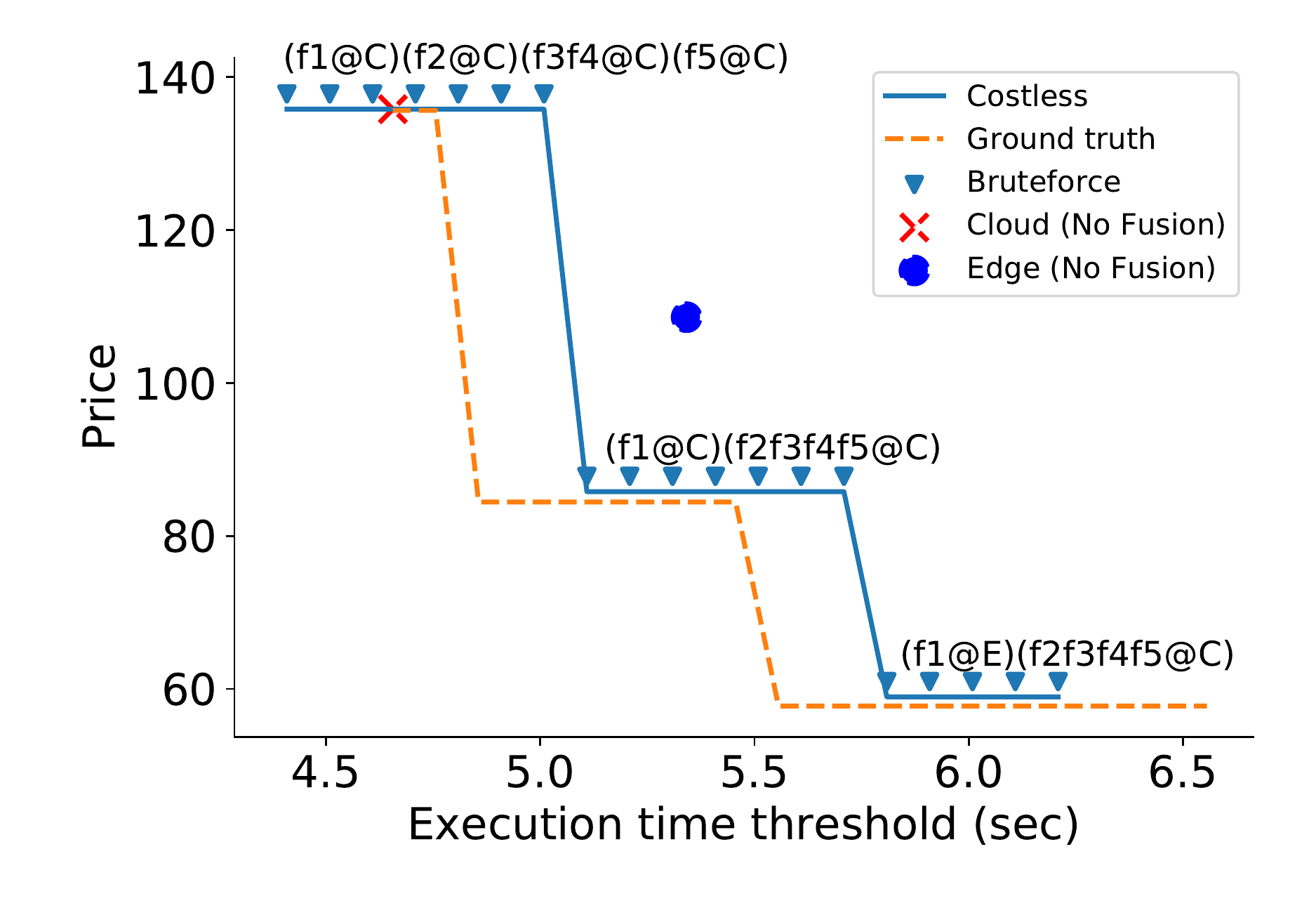}
\caption{Best price below an execution time threshold.}
\label{fig:threshold_2}
\end{figure}

Figure~\ref{fig:threshold_2} shows the results for finding the price for each execution time threshold. The results shows that even though Costless uses an approximate algorithm to solve the constrained shortest path problem, the solutions it found exactly matches the solutions found by Brute force. We also note that the solutions found by Costless are close to the Ground truth obtained from AWS logs, except that Costless sometimes switches to a different price slightly later than the ground truth. We attribute this to the same reason we described in Figure~\ref{fig:time-model} in which the time estimate of Costless is slightly different than the ground truth. However, we can see that Costless eventually reaches the same price 
values obtained from AWS logs and the places where a mismatch occurs is only within 200-300ms.
We further show that the simple policies: {\it Cloud (No Fusion)}, {\it Edge (No fusion)} misses the opportunity to reduce the price with a small difference in execution time, for example, Costless can reduce the cost by 37\% (135\$ - 85\$) with only 5\% increase in latency. The best price Costless can achieve is 58\$ which is 57\% reduction with 15\% increase in latency.

\subsection{Effect of optimizing over memory configurations}
\begin{figure}
\centering
\includegraphics[width=0.5\textwidth,height=0.32\textwidth]{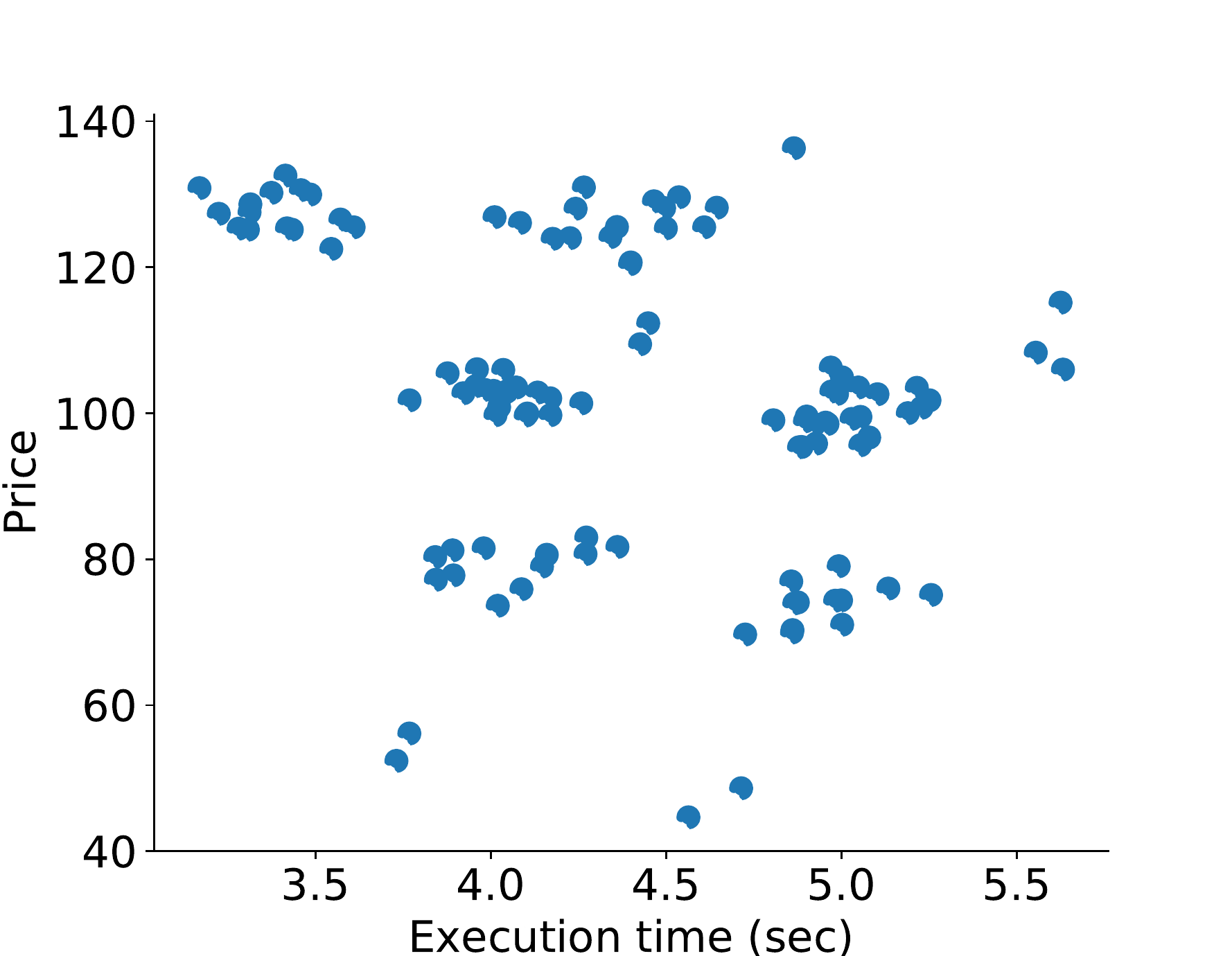}
\caption{Price and time for different fusions and memory configurations }
\label{fig:memory_conf}
\end{figure}

In this section, we show how changing the memory configuration can optimize the price. 
Using the profiling information in Table~\ref{tab:profile} we were able to efficiently search over all the combinations of fusion and memory configurations the same way we search over placement of functions on edge and cloud.  
For each fusion solution, we search over two different memory configurations (128 MB and 256 MB) for each function. We note that for the non-fused case, we have 5 functions so there are $25$ combinations but after fusing we might end up with 2 functions so we search for 4 combinations. Figure~\ref{fig:memory_conf} shows the relationship between the execution time and the price for different solutions of fusion and memory configuration. The Figure shows that the data points are organized into clusters. Each cluster represents one fusion solution and within each cluster there are several data points that have different prices and execution times. We note that for the non-fused case, the solution that have the best price was non-trivial since it keeps the memory for the 4 functions at $128MB$ and it increases the memory of $f3$ only to $256MB$, this ended up being the best solution because $f3$ experienced the highest speedup when the memory increased from $128$ to $256$ (See Table~\ref{tab:profile}). Such speedup has a positive impact on the overall price and execution time. We conclude that setting the memory configurations manually is not ideal and running all profiling configurations is very expensive. On the other hand, profiling each function separately and running a scalable algorithm such as Costless can help finding a non trivial and cheaper memory configurations. The best solution we found improved both the price and the execution time by 6\% and 10\% respectively.

\subsection{Time to find solution}

\begin{figure}
\centering
\includegraphics[width=0.5\textwidth,height=0.35\textwidth]{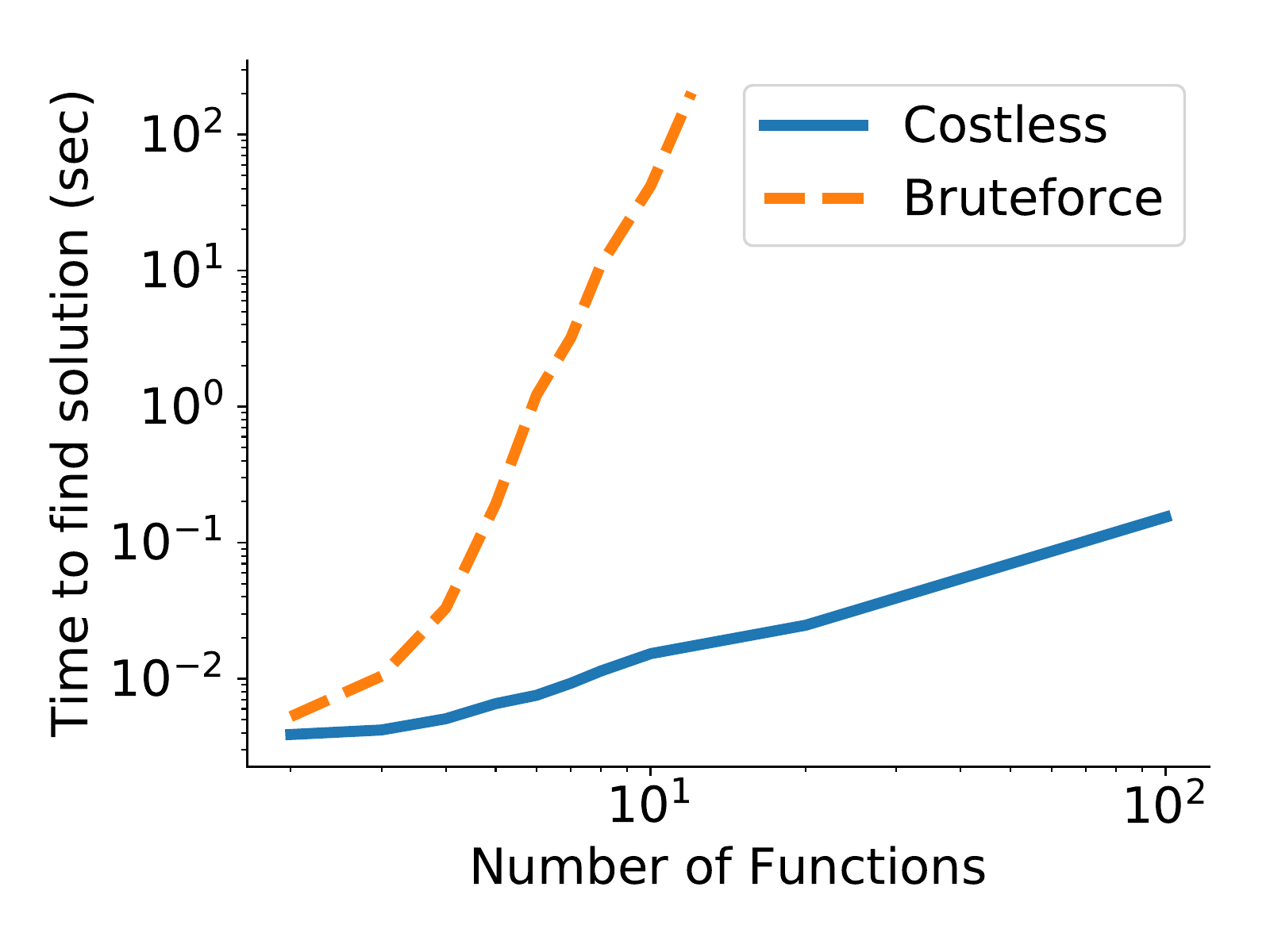}
\caption{Comparison between Costless and brute force in the time in the time to search for the best solution with increasing number of functions}
\label{fig:scalability}
\end{figure}

In this section, we evaluate the scalability of Costless with increasing the number of functions. We generate synthetic functions and we append it to the end of the graph. For each new function we randomly sample its profile which include (1) execution time on the cloud (500s-2s), (2) execution time on edge (1s-5s), and (3) scheduling delay (50ms-300ms).
Figure~\ref{fig:scalability} compares Costless with Bruteforce search, in the time to obtain the placement. The time for Costless includes the construction time of placement graph and executing constrained shortest path algorithm. The brute force time includes calculating the cost for each feasible solution, sorting them based on the price, and finding the lowest price that have latency above certain threshold. With Bruteforce, we were not able to complete the search in a reasonable amount of time when number of functions increased beyond 12. Costless however was able to complete the search within one seconds for 100 functions. The scalability of Costless comes from two main factors: (1) Constructing the cost graph which avoids redundantly computing the cost of fused functions that are shared between multiple solutions, (2) Formulating the problem as a constrained shortest path problem which allows Costless to use scalable heuristics.

%% file: related.tex
\section{Related Work:}\label{sec:related}

{\bf Optimizing cloud computing cost:} There is a significant amount of related work in optimizing the monetary cost charged by public cloud providers while meeting a service level agreement (SLA). The problem has been addressed in the context of Resource provisioning~\cite{perf}
Autoscaling~\cite{autoscale1}\cite{autoscale2} to handle the fluctuations in the user request rate~\cite{elastic1}, short-term on-demand vs long-term reservation plans~\cite{longterm}, and cloud scheduling using spot instances~\cite{risk}\cite{spot}. There are a variety of pricing models addressed in the related work which range from long-term yearly resource reservation plan, to on-demand pay-as-you-go VM instances, to variable-priced VM instances that allows clients to bid for spare CPU-hour resources (e.g., Spot instances). However, such pricing models address the payment for virtual machine usage at a per-hour resolution (as with AWS EC2). To the best of our knowledge this is the first paper to study the factors affecting the pricing of serverless applications and to propose an algorithm to optimize it. In contrary to the classic pricing models that requires per virtual machine hour payment, serverless computing relieves the users from VM provisioning and management, and it allows users to deploy code {\it functions}. This comes with different challenges such as function placement and function fusion. Such problems have not been addressed in the context of serverless computing. However, there exist related work in different research areas that we will summarize next.

{\bf Function Placement} Function placement has been studied in various research areas such: Mobile Computing~\cite{cloudlet}, Service composition~\cite{spidernet}, and Sensor networks~\cite{decentralized-1}. For mobile computing, the concept of cloudlet~\cite{cloudlet} has been proposed as an additional layer that sits between the smartphone and the cloud to provide processing capacity closer to the data. Several heuristics have been proposed to split the processing between the smartphone, cloudlets and clouds to help reduce latencies~\cite{hermes}\cite{genetic}. Service composition is another related problem in which service providers decide on which computing resource the service should be allocated to meet QoS (Quality of service) requirements. The services in this case are similar to the serverless functions. Previous work models service composition as multipath constrained path finding problem~\cite{spidernet}. The fundamental differentiating aspect in our work is the ability to jointly model both the placement and fusion solution in one graph which allows us to model the problem as constrained shortest path and use an efficient algorithms to solve it. 

{\bf Function Fusion} The problem of operator fusion and code generation have received attention in the database systems and high performance computing (HPC) literature because it has the potential to reduce the intermediate data between operators/functions and the number of scans on the input data. However most of the work in these areas deals with a finite set of operators and assumes that the operator semantics are known beforehand. For example the database community deals with relational algebra operators such as Joins, aggregations, and projection~\cite{sparksql}. On the other hand HPC and machine learning communities deals with linear algebra operations such as matrix multiplications and factorizations. Spoof~\cite{spoof} and Tensorflow XLA~\cite{xla} are representative operator fusion approaches in this category, and they focus on searching for patterns of operators that are known to give better performance when fused together. In contrary, our approach depends on profiling the application and is agnostic of the application semantics which is essential given the huge variety of applications that run on edge and cloud platforms. We further explore both placement and fusion solutions which was not addressed in the previous approaches.

%% file: conc.tex
\section{Conclusion}\label{sec:conc}

In this paper, we studied the problem of optimizing the price and execution time for serverless computing. We identified three fundamental factors affecting the price of serverless applications which are: function fusion, function placement, and memory configuration of serverless functions. Our fundamental idea is to represent fusion and placement solutions in one cost graph and propose an efficient algorithm to obtain the best solution given latency or price constraints. Although function fusion has the disadvantage of making the application less modular and maintainable, we show that it is an effective way to reduce the cost, especially when transition cost dominates the function execution cost and we were able to reduce the price of an image processing application by more than 37\% with 5\% increase in the latency and we showed that placement of functions on edge devices can help increase the price reduction to 57\% . We also showed that using the right memory configuration can help reduce both price and latency of the application deployment. We hope that this paper inspires more research in automatic code generation of fused functions and dynamically switching between fused and non-fused deployments.

\section{Acknowledgements}\label{sec:ack}
This work is supported by the National Science Foundation under grants NSF ACI 1659293, NSF ACI 1443013, NSF CCF 14-38982, and NSF CCF 16-17401.